\documentclass[prl,twocolumn,showpacs,preprintnumbers,amsmath,amssymb]{revtex4}
\usepackage{graphicx}
\usepackage{dcolumn}
\usepackage{bm}
\usepackage{trfsigns}
\newcommand \be{\begin{eqnarray}}
\newcommand \ee{\end{eqnarray}}
\newcommand \ba{\begin{align}}

\begin{document}
\title{Reply to the comment of P. Lipavsky}
\author{K. Morawetz$^{1,2,3}$
}
\affiliation{$^1$M\"unster University of Applied Sciences,
Stegerwaldstrasse 39, 48565 Steinfurt, Germany}
\affiliation{$^2$International Institute of Physics- UFRN,
Campus Universit\'ario Lagoa nova,
59078-970 Natal, Brazil}
\affiliation{$^{3}$ Max-Planck-Institute for the Physics of Complex Systems, 01187 Dresden, Germany
}
\begin{abstract}
The critics of P. Lipavsky on the derivation of conservation laws including gradient corrections is refuted since his counterexample is based on a mathematical error. Instead, the derived conservation laws for density, momentum and energy remain valid even without using the objected relation. Only for the entropy local terms are missing which vanish under integration.
\end{abstract}

\maketitle

The comment \cite{L17} concerns the missing power feed due to time-dependent external potentials and the claim that a mathematical identity is wrongly used.

First, the paper \cite{M17} does not include external potentials which lead to power feeds in balance equations. It was not the subject of the paper and a corresponding proof was removed from a former draft of the paper following a mutual agreement. The objected sentence was not intended to make a contrary impression.

Second, the objected mathematical identity (46) of \cite{M17} is valid under integration. The given counter example in the comment does not concern this relations since it contains an error. In the step from (6) to (7) of the comment \cite{L17}, it is assumed wrongly that $\delta(x) f(x)=0$ implies $f(x)=0$ for Dirac-$\delta$ functions. Therefore the argument does not apply and the stipulated assumptions have never been used in the paper. The comment misses also the linear expansion in $\Delta$.

To avoid any misunderstanding, let us repeat the objected steps leading to (46) of \cite{M17}.
We consider $D[\Delta(E)]=f[\Delta(E)]\delta(E-E')$ with a well defined function $f(x)$ and the Dirac-$\delta$ function contained in $D$. 
Firstly, we expand the equality $D[\Delta(E)]=D[\Delta(E')]$ in first order of $\Delta$, 
\be
D(0)+(\partial D) \Delta (E)=D(0)+(\partial D) \Delta (E')
\label{1}
\ee
with any (partial) derivative $\partial$ of a variable.
Secondly, under integration the Dirac $\delta$-function in $D$ leads to
\be
\partial [D\Delta(E)]=0=\partial [D\Delta(E')]
\label{2}
\ee
as seen from partial integration $\partial (f \delta)=f'\delta+f\delta'=f'\delta-f'\delta=0$.
Subtracting the zero (\ref{2}) from (\ref{1}), the desired relation (46)  of \cite{M17},
\be
D \partial \Delta(E)=D \partial \Delta(E'),
\label{relt}
\ee
appears.

A correct argument would have been to observe that this relation holds only under the assumption that we {\em integrate} over the corresponding variable. This is true for the momentum variable, where relation (\ref{relt}) is used when going from (67) to (68) in \cite{M17}. Using it for the time variable when going from (49) to (50) in \cite{M17},  it is not quite obvious that one is not loosing local terms which nevertheless vanish under time integration.

Therefore, it is instructive to see that one obtains exactly the same results without using relation (\ref{relt}). Not employing (\ref{relt}) we have in (50) of \cite{M17} the additional term 
\be
K={\xi_1+\xi_2\over 4} D[\partial_t \Delta_t(E)-\partial_t \Delta_t(E')]
\label{3}
\ee
which we show now to be zero. Employing transformation B which is the interchange of final and initial coordinates, we have $D=2\pi |t|^2\delta (E-E')f_1f_2(1-f_3-f_4)\to D+N$ where $N=2\pi |t|^2\delta (E-E')\{f_3f_4(1-f_1-f_2)-f_1f_2(1-f_3-f_4)\}$, and
\be
K&=&-{\xi_3+\xi_4\over 4} D[\partial_t \Delta_t(E)-\partial_t \Delta_t(E')]
\nonumber\\
&&-{\xi_3+\xi_4\over 4} N[\partial_t \Delta_t(E)-\partial_t \Delta_t(E')].
\label{4}
\ee
For the density $\xi=1$, momentum $\xi=\vec k$ and energy $\xi=\epsilon$ we can replace $\xi_3+\xi_4=\xi_1+\xi_2$ due to the conservation contained in the $\delta$-function of D. Therefore the first term in (\ref{4}) is just $-K$ which leads from (\ref{4}) to
\be
K=-{\xi_3+\xi_4\over 8} N[\partial_t \Delta_t(E)-\partial_t \Delta_t(E')].
\ee
Using this in (\ref{4}) leads to $K=-K+K/2$ which means $K=0$. Therefore, no additional terms appear for density, momentum, and energy, see also \cite{Kl17}.

For the entropy, the additional term (\ref{3}) leads to an additional gain term and a term of symmetrized collision $\sim N$. Both fit in the scheme of the proof of H-theorem in \cite{M17} such that this proof remains valid. Please note that these possible additional local entropy terms vanish under time integration and the claim of the paper \cite{M17} is to have derived all binary-correlated terms which survive under integration. 

In the case of hard-sphere gases, the new correlated entropy gain and current due to the two-particle form does not vanish contrary to the comment. All cited papers of the comment express the binary part of the correlated entropy in terms of  pair-correlation functions. The direct comparison is not yet worked out since these forms have to be translated first into phase-shift expressions as used in the paper \cite{M17}. The claim that the present forms are in conflict, is premature and not justified. 

Summarizing the critics of the mathematical identity (46) of \cite{M17} is not valid. The external power feed was never subject of the paper and cannot be criticized as not having been considered. Possible missing local terms in the entropy are vanishing under time integration and should be worked out as well as the comparison with the literature as it was stated in the paper that this is devoted to forthcoming work.

At the end it should be said that Physical Review E decided not to publish this reply.


\end{document}